\documentclass[twoside,twocolumn,9pt]{article}
\usepackage{extsizes}
\usepackage[super,sort&compress,comma]{natbib} 
\usepackage[version=3]{mhchem}
\usepackage[left=1.5cm, right=1.5cm, top=1.785cm, bottom=2.0cm]{geometry}
\usepackage{balance}
\usepackage{mathptmx}
\usepackage{sectsty}
\usepackage{graphicx} 
\usepackage{lastpage}
\usepackage[format=plain,justification=justified,singlelinecheck=false,font={stretch=1.125,small,sf},labelfont=bf,labelsep=space]{caption}
\usepackage{float}
\usepackage{fancyhdr}
\usepackage{fnpos}
\usepackage{siunitx}

\usepackage[english]{babel}
\addto{\captionsenglish}{%
  
}
\usepackage{array}
\usepackage{droidsans}
\usepackage{charter}
\usepackage[T1]{fontenc}
\usepackage[usenames,dvipsnames]{xcolor}
\usepackage{setspace}
\usepackage[compact]{titlesec}
\usepackage{hyperref}
\usepackage{bm}
\usepackage{amsmath,amssymb}
\usepackage{graphicx}

\usepackage{epstopdf}

\definecolor{cream}{RGB}{222,217,201}

\begin{document}

\pagestyle{fancy}
\thispagestyle{plain}
\fancypagestyle{plain}{
\renewcommand{\headrulewidth}{0pt}
}

\newcommand{\rr}{\bm{r}}
\newcommand{\ii}{\textrm{i}}
\newcommand{\ee}{\textrm{e}}
\newcommand{\dd}{\textrm{d}}
\newcommand{\bigO}{\textrm{O}}
\newcommand{\rad}{\textrm{rad}}
\newcommand{\res}{\textrm{Re}}
\newcommand{\ims}{\textrm{Im}}

\renewcommand{\vec}[1]{\bm{#1}}

\makeFNbottom
\makeatletter
\renewcommand\LARGE{\@setfontsize\LARGE{15pt}{17}}
\renewcommand\Large{\@setfontsize\Large{12pt}{14}}
\renewcommand\large{\@setfontsize\large{10pt}{12}}
\renewcommand\footnotesize{\@setfontsize\footnotesize{7pt}{10}}
\makeatother

\renewcommand{\thefootnote}{\fnsymbol{footnote}}
\renewcommand\footnoterule{\vspace*{1pt}%
\color{cream}\hrule width 3.5in height 0.4pt \color{black}\vspace*{5pt}} 
\setcounter{secnumdepth}{5}

\makeatletter 
\renewcommand\@biblabel[1]{#1}            
\renewcommand\@makefntext[1]%
{\noindent\makebox[0pt][r]{\@thefnmark\,}#1}
\makeatother 
\renewcommand{\figurename}{\small{Fig.}~}
\sectionfont{\sffamily\Large}
\subsectionfont{\normalsize}
\subsubsectionfont{\bf}
\setstretch{1.125} 
\setlength{\skip\footins}{0.8cm}
\setlength{\footnotesep}{0.25cm}
\setlength{\jot}{10pt}
\titlespacing*{\section}{0pt}{4pt}{4pt}
\titlespacing*{\subsection}{0pt}{15pt}{1pt}

\fancyfoot{}
\fancyfoot[RO]{\footnotesize{\sffamily{1--\pageref{LastPage} ~\textbar  \hspace{2pt}\thepage}}}
\fancyfoot[LE]{\footnotesize{\sffamily{\thepage~\textbar\hspace{3.45cm} 1--\pageref{LastPage}}}}
\fancyhead{}
\renewcommand{\headrulewidth}{0pt} 
\renewcommand{\footrulewidth}{0pt}
\setlength{\arrayrulewidth}{1pt}
\setlength{\columnsep}{6.5mm}
\setlength\bibsep{1pt}

\makeatletter 
\newlength{\figrulesep} 
\setlength{\figrulesep}{0.5\textfloatsep} 

\newcommand{\topfigrule}{\vspace*{-1pt}%
\noindent{\color{cream}\rule[-\figrulesep]{\columnwidth}{1.5pt}} }

\newcommand{\botfigrule}{\vspace*{-2pt}%
\noindent{\color{cream}\rule[\figrulesep]{\columnwidth}{1.5pt}} }

\newcommand{\dblfigrule}{\vspace*{-1pt}%
\noindent{\color{cream}\rule[-\figrulesep]{\textwidth}{1.5pt}} }

\makeatother

\twocolumn[
  \begin{@twocolumnfalse}
\sffamily
 \LARGE{
 \noindent
 \textbf{Raman-acoustofluidic integrated system for single-cell analysis$^\dag$}
 } \\
\\
\large{
     \noindent
     Harrisson D. A. Santos,$^{\ast\ddag}$\textit{$^{a}$}
     Amanda E. Silva,$^\ast$\textit{$^{b}$}
     Gicl\^enio C. Silva,$^\ast$\textit{$^{a}$}
     Everton B. Lima,\textit{$^{a}$}
     Alisson S. Marques,\textit{$^{a}$} Magna S. Alexandre-Moreira,\textit{$^{b}$}
     Aline C. Queiroz,\textit{$^{c}$}
     Carlos Jacinto,\textit{$^{d}$} 
     J. Henrique Lopes,\textit{$^{e}$} 
     Ueslen Rocha,\textit{$^{d}$} and 
     Glauber T. Silva$^{\ddag}$\textit{$^{a}$}
} \\
  
\normalsize{
    \noindent
    Confocal Raman microscopy offers a particular pathway for monitoring chemical ``fingerprints'' of intracellular components like the cell membrane, organelles, and nucleus. 
    Nevertheless, conventional Raman acquisitions of fixed or randomly dispersed cells on a substrate, such as a microscope slide, might be severely limited for biomedical investigations.
    Backscattered Raman signal is likely to be overshadowed by contributions from substrate fluorescence.
    Also, biological assays for drug discovery, infection, and tissue engineering may require monitoring live cells individually over a time interval spanning from a few hours to a few days.
    To meet the needs of cell assay monitoring, we propose an acoustofluidic device  that forms a levitating cell aggregation (one layer) in a  cylindrical chamber of  $\SI{10}{\micro\liter}$ volume that operates at $\SI{1}{\mega\hertz}$ frequency.
    The  integrated system comprises a lab-on-a-chip device and a confocal Raman microscope. 
    In this setup, a cell can be selectively Raman-investigated with micrometre accuracy, without any substrate interference. 
    Based on a set of carefully designed experiments, we demonstrate that polystyrene microparticles are assembled and held standstill, enabling a full Raman spectrum to be taken of a single particle in less than a minute.
    The signal-to-background ratio is improved by a thousandfold when compared with conventional Raman acquisition. 
    The Raman-acoustofluidic system is showcased for obtaining the spectrum of a single cell among an enriched population of macrophages of mice.
    The obtained results confirm the method's robustness for applications in single-cell analysis.
} 
\\
\vspace{0.6cm}
\end{@twocolumnfalse} 
]


\renewcommand*\rmdefault{bch}\normalfont\upshape
\rmfamily
\section*{}
\vspace{-1cm}


\footnotetext{
\textit{$^{a}$~Physical Acoustics Group, Instituto de Fisica, Universidade Federal de Alagoas,}}
\footnotetext{\textit{$^{b}$~Laborat\'orio de Farmacologia e Imunidade, Instituto de Ci\^encias Biol\'ogicas e da Sa\'ude, Universidade Federal de Alagoas,}}
\footnotetext{\textit{$^{b}$~Centro de Ci\^encias M\'edicas e de Enfermagem, Universidade Federal de Alagoas (Campus Arapiraca),}}
\footnotetext{\textit{$^{d}$~Grupo de Nano-Fotônica e Imagens, Instituto de F\'isica, Universidade Federal de Alagoas,}}
\footnotetext{\textit{$^{e}$~Grupo de Ac\'ustica e Aplica\c{c}\~oes, N\'ucleo de Ci\^encias Exatas, Universidade Federal de Alagoas (Campus Arapiraca).}}
\footnotetext{$\ast$~These authors contributed equally to this work.}
\footnotetext{\ddag~Corresponding authors. E-mail: hdasantos@fis.ufal.br, gtomaz@fis.ufal.br}
\footnotetext{\dag~Electronic Supplementary Information (ESI) available.}


\section*{Introduction}
Raman spectroscopy is a well-known technique widely used in the last decade for cell and tissue diagnosis, and therapy purposes.\cite{seo2014nir,Cui2018}
Bioevents at the cellular level can be assessed from the Raman spectrum related to molecules from cell membranes, organelles, proteins, ribosomes, lipids, and DNA. 
Notable investigations include benign and malignant cell lines,\cite{Taleb_2006, Chan2009} randomly distributed cell recognition,\cite{Notingher2004,Krishna2005}
neoplastic and normal hematopoietic cells;\cite{Chan2006} cancer diagnosis,\cite{Frank1995,Krishna2004} cellular
proliferation,\cite{Short2005,Alraies2019} and drug trials.\cite{Zivanovic2018,Bik2019}

Cell assays are preferably performed in an aqueous environment in order to mimic normal physiological conditions.\cite{Nguyen2017} 
In such configuration, Raman spectroscopy usually yields superior results as compared to infrared (IR) and fluorescence spectroscopy because it avoids interference from water-like solutions.\cite{Li2014} 
Although Raman spectroscopy can assess cellular status, the small probability of a Raman  event\cite{Zhao2008,Jones2019} alongside the corresponding small scattering cross-section  of most biological macromolecules\cite{PUPPELS_1991} contribute to a low signal-to-background ratio. 
To further complicate matters, randomly dispersed cells may also reduce the probability of scattering events. 
This problem can be circumvented by fixing  cells  using organic substances at the expense of inducing misinformation in the  spectrum of cellular structures.\cite{Mason1991,Li2017}

An alternative to fixture methods or
dispersed cells is the Raman-optical tweezer technique,\cite{Snook2009} where live cells can be analysed one at the time.
Optical tweezers are based on the radiation pressure of light that is proportional to the field intensity per speed of light. 
In turn, the large value of the speed of light imposes the application of high-intensity fields to achieve proper tweezing forces on cells. 
As a consequence, photothermal and photochemical damages may occur, affecting cellular processes.\cite{BlazquezCastro2019} 
Cell trapping can also be achieved by means of the acoustic radiation force of ultrasonic waves, which generally employs less power than optical tweezers.\cite{Dholakia2020}  
The acoustic radiation force has been systematically harnessed in microfluidic lab-on-a-chip devices for   microparticle handling in the so-called acoustofluidics realm.\cite{Bruus2011} 
In a typical arrangement, microparticles are  injected into a small-volume chamber (with a few microlitres or less) and 
the acoustic radiation force aggregates them  in a levitating state.\cite{Spengler2001} 
Cell cultures have been maintained in acoustic trapping conditions for  as long as seven days.\cite{Christakou2015} 
Thus, acoustofluidics offers an outstanding framework to mimic the living-cell conditions, enabling a highly selective investigation through Raman spectroscopy. 

A method based on the acoustic levitation of droplets  in the air has been used for Raman investigations.\cite{Biswas1995,Santesson2003,Leopold2003,LopezPastor2007,Quino2015,Wood2005,Puskar2007}
Even so this method provides a stable noncontact approach, it cannot hold the sample over a long time due to droplet evaporation.
On the other hand,  an acoustofluidic device with trapped microparticles in a liquid flow was used for Raman spectroscopy of chemical reactions.\cite{RuedasRama2007, Wieland2019} 
Recently,  Raman-acoustofluidic setups were proposed for real-time assessment of live mycobacteria\cite{Baron2020} 
and fingerprint detection at low DNA concentrations.\cite{Xu2020}
Despite the excellent results obtained so far, a systematic and detailed study about quantitative improvements offered by acoustofluidics for Raman spectroscopy is yet to be reported.
The method's relevant aspects for  single-cell investigation need also to be addressed.

In this work, we present a timely Raman-acoustofluidic integrated system for single-cell monitoring. 
The  system is composed of a  3D-printed lab-on-a-chip device with a cylindrical chamber of $\SI{10}{\micro\liter}$ volume filled with an aqueous solution whereby the microconstituents are dispersed in; and a commercial confocal Raman microscope.
The chamber is sealed at the top and bottom by a glass slide and a piezoelectric emitter that operates in a low-voltage regime ($<\SI{10}{\volt}$) at $\SI{1}{\mega\hertz}$ frequency generating a half-wavelength axial resonator.  
With the help of the acoustic radiation force, the microparticles are vertically trapped at the height of nearly $\SI{285}{\micro\meter}$ with a single-layer thickness.
The one-layer feature along with the high-resolution Raman system\cite{Corle1996} ($\SI{1.5}{\micro\meter}$ transversely and  $\SI{2.0}{\micro\meter}$ in depth) allow an individual particle analysis.
%
By keeping a low-voltage supply, the microparticle cluster is standstill for several hours without increasing the chamber temperature.
The spectrum of $10$ and $\SI{30}{\micro\meter}$-diameter polystyrene particles is obtained with a $40\times$ objective lens. 
The  signal-to-background noise is remarkably improved by a thousandfold thanks to the nonphysical contact between the microparticles with the chamber solid parts.
Also, system features such as trap stability, particle concentration, temperature rise, inertial cavitation, and cell viability are discussed in detail for the proposed method.
Finally, our method is also showcased for obtaining the spectrum of live macrophages of mice with no special preparation. 
We achieved a high-selectivity with the spectrum corresponding to a single-cell investigation.
Additionally, our results are remarkably consistent with a previously obtained spectrum from fixed macrophages on a silicon wafer substrate.\cite{Araujo2019}
\begin{figure*}
    \begin{center}
        \includegraphics[scale=2.7]{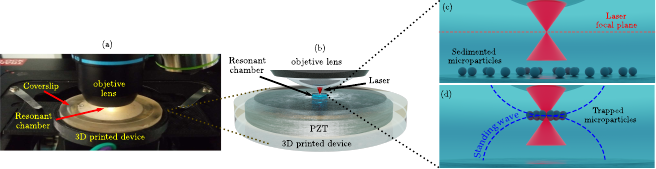}
        \caption{
            (a) Photography of the Raman-acoustofluidic integrated system comprises a cylindrical acoustic chamber and a confocal Raman apparatus.
            (b) Zoomed-in illustration of the 3D printed acoustofluidic device with the resonant chamber ($\SI{40}{\milli\meter}$ diameter and $\SI{750}{\micro\meter}$ height) in light blue. 
            A piezo-ceramic actuator (PZT) is glued underneath the chamber, and a glass slide is placed on its top.
            (c) Microparticles sedimented at the chamber bottom.
            The focused Raman laser is depicted as red cones.
            (d) After switching on the device, microparticles are trapped in the central area of the nodal pressure plane. 
            The blue-dashed curve depicts the acoustic standing wave in the axial direction.
           \label{fig:acoustofluidic-Raman}
        }
    \end{center}
\end{figure*}

\section*{Experimental setup}

\subsection*{Acoustofluidic device fabrication}
The acoustic trapping of microparticles or cells are performed 
in a cylindrical acoustic chamber with a height of $H=\SI{750}{\micro\meter}$ and
diameter of $2R=\SI{4}{\milli\meter}$.
The chamber is cast inside a cylindrical disk, which is fabricated with a 3D printer (Moonray D75, Sprintray, Inc., USA) through the digital light processing technique. 
A piezoceramic actuator (lead zirconate titanate, PZT-8) with a diameter of $\SI{25}{\milli\meter}$ is glued underneath the resonant chamber with epoxy (Huntsman, Corp., USA).
A glass cover slide of a $\SI{150}{\micro\meter}$ thickness is placed at the chamber's top, working as an acoustic reflector.
In Fig.~\ref{fig:acoustofluidic-Raman}, we illustrate the acoustofluidic device.

\subsection*{Raman spectroscopy}
The Raman measurements are carried out by  a confocal Raman microscope (LabRam HR Evolution, HORIBA, France).
The microscope comprises  a $40\times$ (with a numerical aperture $\text{NA} = 0.65$) objective lens and a CW diode laser
which generates a $\SI{785}{\nano\meter}$-laser with a power under $\SI{100}{\milli\watt}$. 
The Raman backscattered signal is collected by the same lens and is dispersed by $300$\, grooves/\si{\milli\meter}. 
The system is calibrated using the silicon phonon band at $\SI{520}{\per\centi\meter}$ as a reference. 
The Raman laser is positioned by a controllable $xyz$ translational stage.

\subsection*{Electronic instrumentation}
The piezoceramic actuator of the acoustofluidic device is excited with a sinusoidal signal produced by a function generator (AFG1022, Tektronix, Inc., USA) and amplified by an RF power amplifier (240L, Electronics \& Innovation, Ltd., USA).
The driving signals are monitored with a two-channel oscilloscope (TDS 2012C, Tektronix, Inc., USA).

\subsection*{Cell preparation}
The adherent-phenotype macrophage line j774.A1 is cultured in Dulbecco's Modified Eagle's Medium (DMEM, Merck KGaA, Germany) supplemented with $10\%$ FBS at \SI{37}{\celsius}, $95\%$ humidity and $5\%$ CO$_2$. 
Cells are cultured to $90\%$ confluence and later centrifuged at $1500$\,rpm for $\SI{5}{\minute}$ at $\SI{4}{\celsius}$ to separate dead cells from living cells. 
Subsequently, the living cells are counted using an optical microscope and a Neubauer's chamber.

\section*{Results and discussion}

\subsection*{Raman-acoustofluidic integrated system}
In Fig.~\ref{fig:acoustofluidic-Raman}, we depict the Raman-acoustofludic platform proposed for cell enrichment and enhancement of the  Raman signal to background ratio.
A photography of the acoustofluidic device mounted on the confocal Raman microscope is shown in panel (a).
While in panel (b), we illustrate the acoustofluidic device's schematics.
Ultrasonic waves at $\SI{1.056}{\mega\hertz}$ frequency are pumped into the resonant chamber by the piezoceramic actuator. 
A glass coverslip works as an acoustic reflector, which forms a half-wavelength resonant chamber.
Thus,  microparticles are expected to levitate in a pressure nodal plane of height $h\sim\lambda_\text{ac}/2=\SI{325}{\micro\meter}$.
Acoustic levitation is essential to allow the visualization/investigation of the microparticles through confocal microscopy by matching the acoustic trapping plane to
the confocal plane of the objective lens.
Hence, the Raman-acoustofludic system should have a lens working distance satisfying
\begin{equation}
\text{WD}\ge
    \delta_\text{ref} +
    h,
\end{equation}
where $\delta_\text{ref}$ is 
the thickness of the acoustic reflector.
The working distance for our system reads
$\text{WD}\ge \SI{475}{\micro\meter}$.
This gives us reasons to use  a $40\times$ objective lens, with $\text{NA}=0.65$ and a  working distance of $\SI{600}{\micro\meter}$,
to focus the Raman laser in the levitation plane (see Fig.~\ref{fig:acoustofluidic-Raman}).

\subsection*{Acoustic fields}
We assume the fluid inside the chamber is characterised by mass density $\rho_0$ and adiabatic compressibility $\beta_0$.
The piezoelectric vibrations at angular frequency $\omega$ induces
a harmonic acoustic pressure and 
fluid velocity whose spatial amplitude is denoted by 
$p_\text{ch}$ and  $\vec{v}_\text{ch}$, respectively.
For applications in the Raman-acoustofluidic system, we seek radially symmetric resonant modes.
In so doing, we use Comsol Multiphysics Software (Comsol Inc., USA) to compute the acoustic fields numerically  
through the finite element method.
The numerical model considers the device  geometry and its material composition (fluid, viscoelastic and elastic solid, and piezoelectric parts) for the acoustic simulations--see details on the the ESI.\dag\:
The  pressure and fluid velocity will be used in the next section to calculate the acoustic radiation force. 

\subsection*{Acoustic radiation force}
The acoustic radiation force exerted on microparticles results from a change in the linear-momentum flux of an incident wave during the scattering process.
In our approach, we assume the microparticles are much smaller than the acoustic wavelength, e.g., $a\ll \lambda_\textrm{ac}$. 
Hence, the primary radiation force is given as minus the gradient  of a potential function,\cite{Bruus2012, Silva2014}
\begin{subequations}
    \begin{align}
    \label{Frad}
        \vec{F}^\rad &= -\nabla U^\text{rad}, \\
        U^\text{rad} &= \frac{4\pi a^3}{3}
            \left( \frac{f_0}{2} \beta_0 |p_\text{ch}|^2 - 
            \frac{3 f_1}{4} \rho_0 |v_\text{ch}|^2
            \right),
    \label{Upotential}
    \end{align}
\end{subequations}
The coefficients $f_0$ and $f_1$ represent the compressibility and density contrast factors between the particle and surrounding liquid.
The minima of the radiation force potential correspond to
the acoustic traps.
%
%
\begin{figure}
    \begin{center}
        \includegraphics[scale=.36]{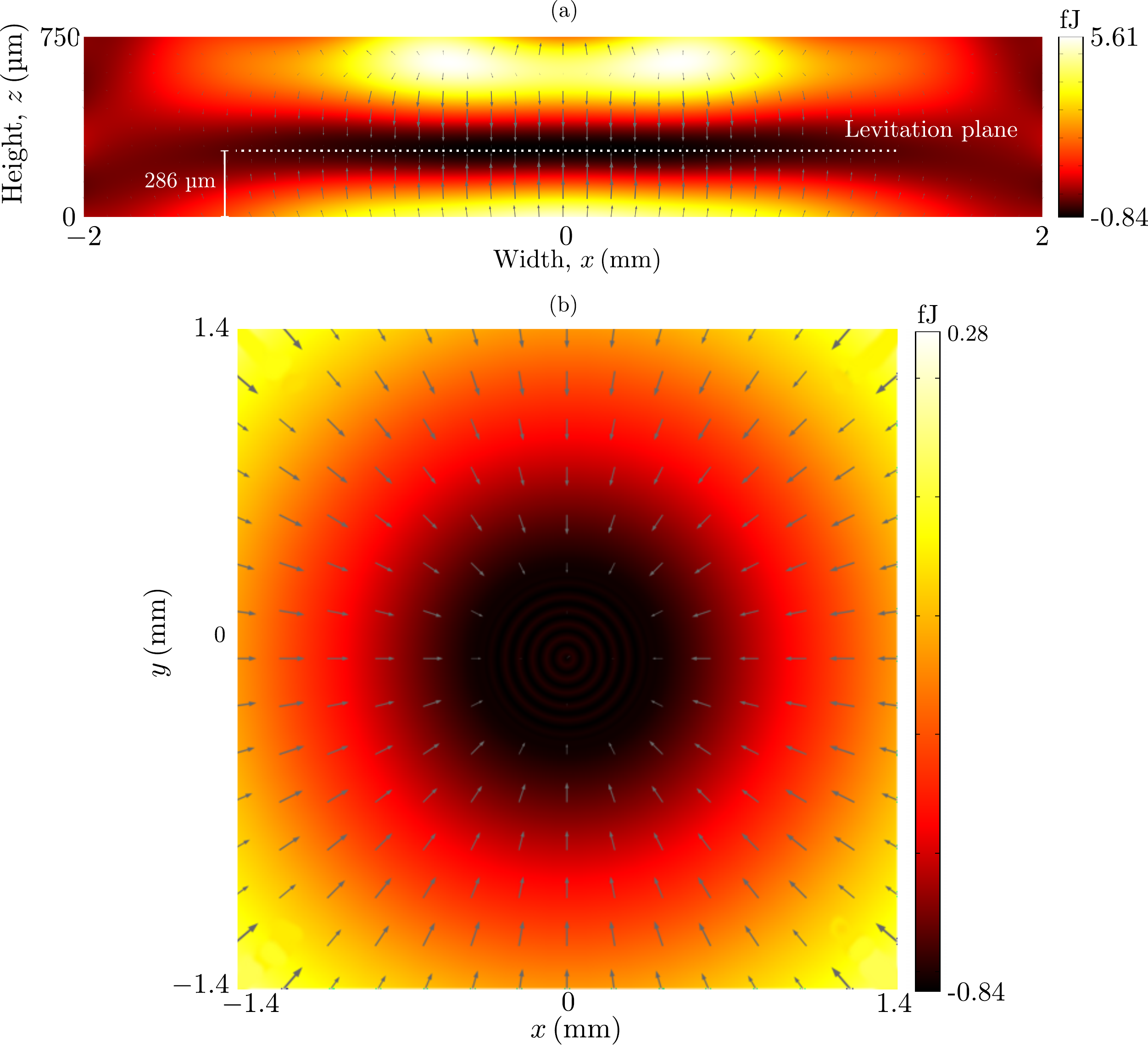}
        \caption{
       The finite element simulation  of the acoustic radiation force field
        represented by grey arrows exerted on a $\SI{10}{\micro\meter}$-polystyrene bead.
        The background images correspond to the radiation force potential.
        The device is simulated with a $\SI{1.19}{\mega\hertz}$ frequency and a peak-to-peak voltage of $\SI{3.8}{\volt}$.
        Panels (a) and (b) show the potential in the axial ($xz$ plane) and levitation plane ($xy$ plane) at $h=\SI{286}{\micro\meter}$, respectively.
        In panel (a),
        the white dotted line depicts the potential minimum position, which corresponds to the levitation plane.
        \label{fig:comsol_simulation}
        }
    \end{center}
\end{figure}

The computed pressure field inside the resonant chamber is used to calculate the acoustic radiation force exerted on a $\SI{10}{\micro\meter}$-polystyrene bead (see details of the computational simulations in the ESI\dag).
The acoustofluidic device is simulated with a frequency of $\SI{1.19}{\mega\hertz}$ and a $\SI{3.8}{\volt}$ peak-to-peak voltage.
In Fig.~\ref{fig:comsol_simulation}, the radiation force potential is illustrated with the origin of the coordinate system set in the principal chamber axis at the bottom.
Panel (a) shows the axial plane's radiation force potential ($xz$ plane). 
Whereas panel (b) depicts the same as (a) but in the levitation plane ($xy$ plane).
The grey arrows represent the radiation force vector field.
In panel (a), the potential minimum is shown in the white dotted line, which corresponds to the levitation plane at  $h=\SI{286}{\micro\meter}$. 
This is in good agreement with the experimentally measured height $h=\SI[separate-uncertainty = true]{285(15)}{\micro\meter}$ at
a frequency of $\SI{1.056}{\mega\hertz}$.
The radiation force action will make particles aggregate in the central portion of the chamber.

The estimation of the ultrasonic wave's pressure amplitude inside the chamber is important in the cell viability analysis.\cite{Wiklund2012a}
To perform this task, we use the gravity-radiation force balance method presented for a particle in a standing plane wave.\cite{Spengler2001} 
Considering a $\SI{10}{\micro\meter}$-polystyrene bead in water at room temperature with a $\SI{1.056}{\mega\hertz}$ frequency, we have the following model parameters
$a=\SI{5}{\micro\meter}$, $f_0=0.31$, $f_1=0.03$, $\rho_0=\SI{998}{\kilogram\per\meter\cubed}$, and $c_0=\SI{1493}{\meter\per\second}$.
Hence, the pressure amplitude is about $p_0=\SI{0.05}{\mega\pascal}$, with a stored acoustic energy density of $E_0=\SI{0.55}{\joule\per\meter\cubed}$.

When two or more microparticles in the nodal plane are in close proximity, the \textit{secondary} radiation force (also known as the acoustic interaction force)  arises on the microparticles due to re-scattering events.\cite{Silva2014a,Lopes2016}
Consider a pair of identical trapped particles, namely particle 1 and 2, with inter-particle distance $d$.
Say particle 1 scatters the incoming wave, with
the scattered wave's fluid velocity $\vec{v}_\textrm{sc}$ being given regarding the reference frame in particle 1.
The acoustic interaction force on particle 2 is\cite{Silva2014a}
\begin{subequations}
    \begin{align}    
    \vec{F}^\text{int} &=
    -\nabla_\perp  U^\text{int},\\
    U^\text{int} &= -\pi a^3 f_1 \rho_0\,
    \text{Re}
        \left[  
         \vec{v}_\text{ch}^*\cdot \vec{v}_\text{sc}
        \right],
    \end{align}
\end{subequations}
where
$\nabla_\perp$ is the transverse gradient, `Re' means the real part of a complex function, and dot denotes the scalar product.

In the levitation plane, we expect the acoustic interaction force to be a central force, i.e., the interaction potential $U^\text{int}$ depends on the inter-particle distance only.
To estimate the acoustic interaction force, we first notice that the nearfield scattered-wave velocity is proportional to\cite{pierce2019acoustics} $\vec{v}_\text{sc} \sim a^3 f_1 \vec{v}_\text{ch}/r^3$, with $r$ being the distance from the center of particle 2 to the observation point.
Hence, the nearfield  interaction potential reads
$U^\text{int}\sim -\pi a^6 f_1^2 \rho_0 |v_\text{ch}|^2/r^3$.
Moreover, the only spatial variations of the 
 acoustic pressure is of the order of the chamber radius $R$,
$\nabla_\perp |v_\text{ch}|^2\sim |v_\text{ch}|^2/R$.
Therefore, we estimate the acoustic interaction force   as
$F^\text{int} = -\nabla_\perp U^\text{int}|_{r=d} \sim \pi a^6 f_1^2 \rho_0 |v_\text{ch}|^2/d^4$.

To compare the interaction force with the primary radiation force (in the nodal plane), we use Eq.~\eqref{Frad} to obtain the ratio
\begin{equation}
    \frac{F^\text{int}}{F^\text{rad}}
    \sim \frac{a^3 R f_1}{d^4}.
\end{equation}
For two $\SI{10}{\micro\meter}$-diameter polystyrene particles in water,
this ratio is $0.75$, with $d=2a$.
We see that in close proximity, the primary and secondary radiation forces are about the same order of magnitude.
It is important to fully determine the secondary radiation force in the levitation plane
as it provides us with a basic understanding of microparticle close-packing arrangements.
However, the aforesaid analysis is beyond the scope of this study.

\subsection*{Acoustic microstreaming}
In addition to transverse radiation forces in the levitation plane, acoustic microstreaming exerts a drag on aggregated particles causing their in-plane movement.\cite{Wiklund2012}
Nonetheless,
Raman spectroscopy should be performed on a standstill particle for as long as the acquisition process requires.
Microstreaming can then be a showstopper of the   Raman-acoustofluidics as a tool for selectively monitoring single cells.
Particles with a diameter smaller than $\SI{15}{\micro\meter}$ are more prone to microstreaming effects.\cite{Spengler2003} 
A possible route to suppress acoustic microstreaming is to develop shape-optimized chambers.\cite{Bach2020} 
Furthermore, the method selectivity may benefit from patterned particles in the one-cell-per-trap configuration.\cite{Collins2015,Silva2019}
\begin{figure}
    \begin{center}
        \includegraphics[scale=.63]{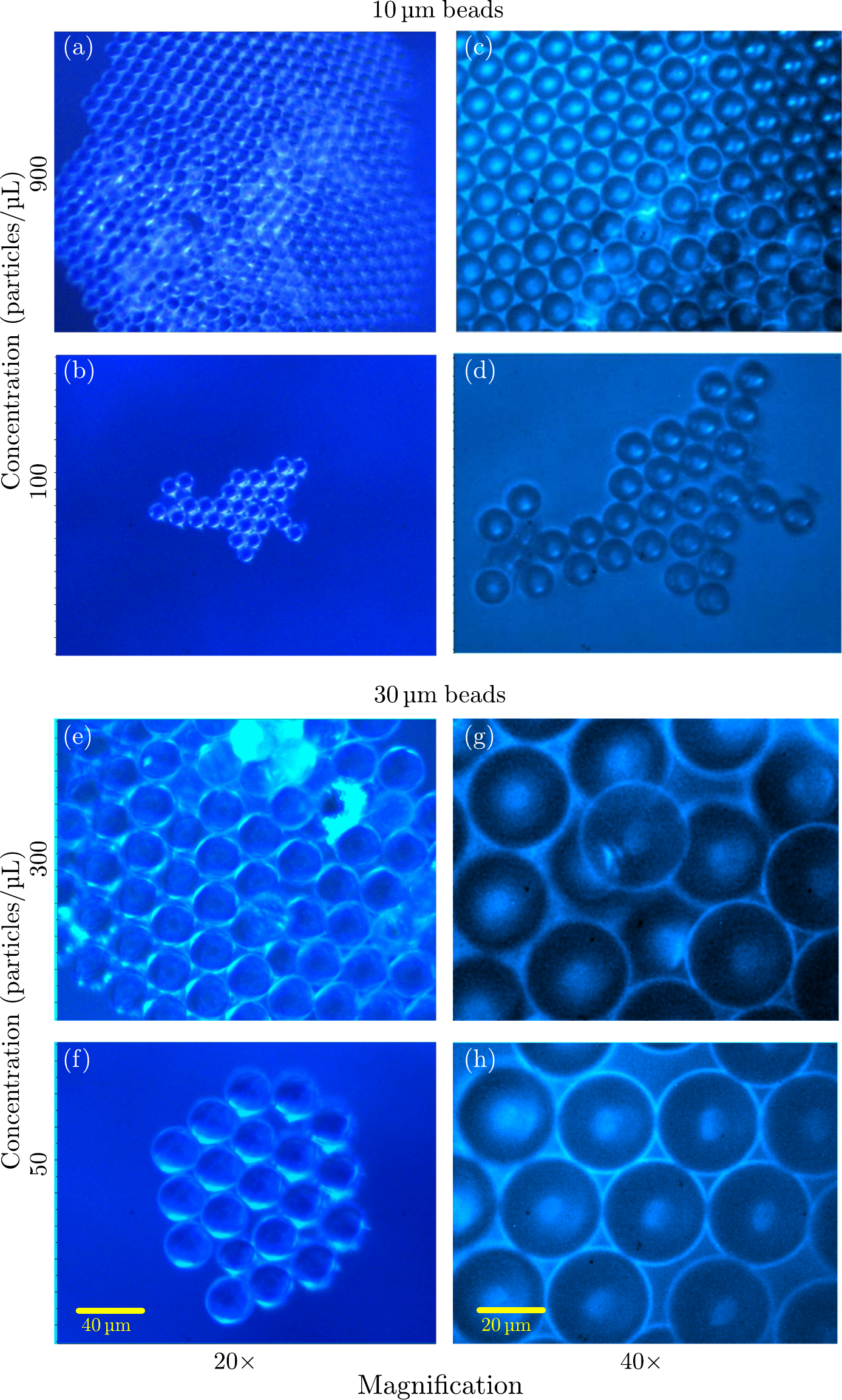}
        \caption{
            Micrographs of monodispersed  polystyrene beads immersed distilled water in the
            acoustofluidic device.
            (a)-(d)  $\SI{10}{\micro\meter}$-diameter beads.
            (e)-(h) $\SI{30}{\micro\meter}$-diameter beads.
            The microparticles are trapped at $\SI{280}{\micro\meter}$ height in the center of the resonant chamber.
            The  device operates at $\SI{1.056}{\mega\hertz}$ with a voltage amplitude of $\SI{3.8}{\volt}$.
            Different  concentrations are illustrated versus optical magnification.
 \label{fig:10poly-particles}
        }
    \end{center}
\end{figure}

\subsection*{Polystyrene beads}
Our first experiment using the Raman-acoustofluidic settings is performed with monodispersed polystyrene beads immersed in distilled water.
Here, the acoustofluidic device operates with $\SI{1.056}{\mega\hertz}$ and a low peak-to-peak voltage of $V_\text{pp}=\SI{3.8}{\volt}$.

In Fig.~\ref{fig:10poly-particles}, we show the micrographs of aggregated polystyrene beads in the device chamber.
Panels (a)-(d) depict $\SI{10}{\micro\meter}$-diameter beads, whereas panels (e)-(h) show
$\SI{30}{\micro\meter}$-diameter particles.
The particle concentration ranges from
$C=50$ to $900\,\textrm{particles/\si{\micro\liter}}$, with a $20\times$ and $40\times$ objective lens (see Figs. S1 and S2 for more details\dag).
The microparticles are trapped in a single layer in less than one minute after switching on the device.
The aggregation efficiency is defined as the ratio between the number of particles $N$ counted in the central area of the levitation plane and the particle concentration $C$ times the chamber's volume $V_\text{ch}$,
\begin{equation}
    \epsilon = \frac{N}{C V_\text{ch}}.
\end{equation}
Clearly, $\epsilon=1$ means all diluted particles are trapped in the levitation plane.
Based on the number of particles seen in panel (b) and (f), the  efficiencies are $\epsilon_{10}=0.28$ and
$\epsilon_{30}=0.40$, for the $\SI{10}{\micro\meter}$ and $\SI{30}{\micro\meter}$ particles, respectively.
The efficiency $\epsilon_{30}$ is larger because the primary radiation force is proportional to the particle cross-section area.

Regarding the close-packed arrangement seen in Fig.~\ref{fig:10poly-particles}, we notice aggregations form a hexagonal lattice.
Computer simulations considering rigid particles demonstrate that a central force leads to hexagonal lattice packing.\cite{Campello2016}
Since Coulomb forces play no role in microparticle interactions here, the only central force present in the nodal pressure plane is the secondary radiation force.
So, we conclude this force causes the observed hexagonal lattice configuration of particles.
As the observed aggregates remain stable for at least a few hours, we could selectively complete Raman acquisitions of any particle in the bead ensemble.
It is also possible to take a Raman map considering different points of the same particle.
With small particle concentrations ($C<50$\,particles/\si{\micro\liter}),  we observed  displacements of a few tens of micrometers of the   $\SI{10}{\micro\meter}$-bead aggregate due to acoustic microstreaming.
Aggregates with larger beads remained standstill.
Furthermore, we noticed that the Raman laser could eject peripheral particles out of an a aggregate. 
Indeed the acoustic trap breakup by light has been experimentally observed.\cite{Dumy2019}
We found that the particle-ejection effect by light is more likely on $\SI{10}{\micro\meter}$ beads with laser intensities greater than $\SI{5}{\milli\watt}$.
\begin{figure}
    \begin{center}
        \includegraphics[scale=.97]{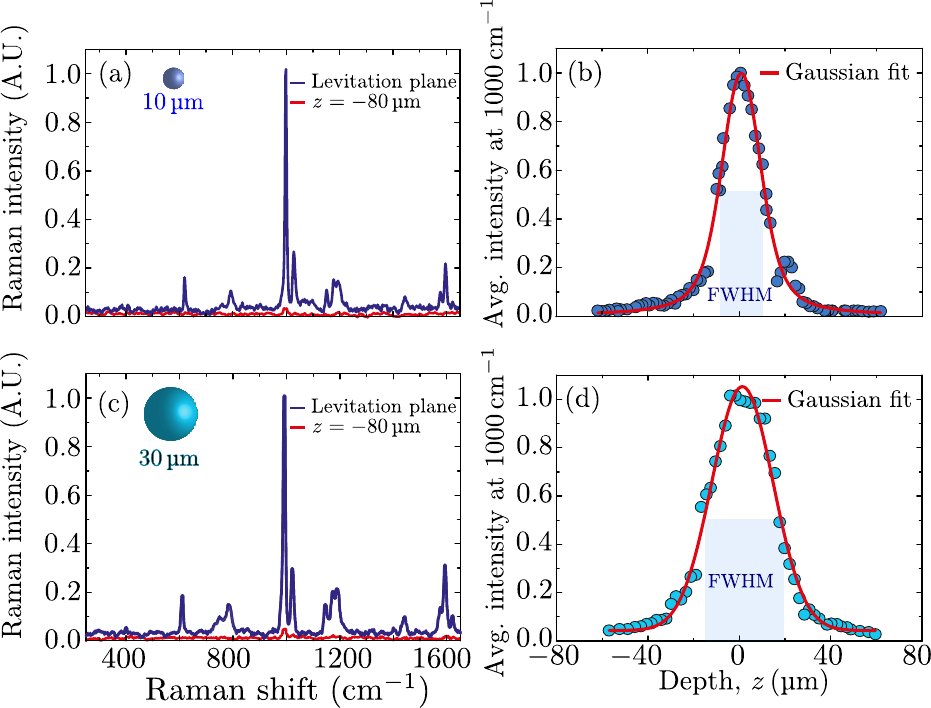}
        \caption{
            The Raman spectrum of (a) a $\SI{10}{\micro\meter}$- and (c) $\SI{30}{\micro\meter}$-diameter polystyrene particle at different depths $z=0$ (levitation plane) and $z=-\SI{80}{\micro\meter}$ inside the resonant chamber.
            The polystyrene particles are dispersed in distilled  water with concentrations of $C=50$ and $100$\,particles/$\si{\micro\liter}$  for sizes of $10$ and $\SI{30}{\micro\meter}$, respectively.
            In both cases, the Raman spectrum is taken by a $\SI{785}{\nano\meter}$ excitation laser of $\SI{2}{\milli\watt}$ power with $\SI{30}{\second}$ acquisition time. 
            The average value around the  $\SI{1000}{\per\centi\meter}$ peak 
            is used to assess the Raman signal axial broadening, as shown in panels (b) and (d). 
            The full width at half maximum (FWHM) is  (b) $\SI{15}{\micro\meter}$ and (d) $\SI{33}{\micro\meter}$, with
            the red line  depicting a Gaussian fit of the measured data.
 \label{fig:Raman-particles}
        }
    \end{center}
\end{figure}

In Fig.~\ref{fig:Raman-particles}, we show the Raman spectrum of (a) a $\SI{10}{\micro\meter}$ and (c) a $\SI{30}{\micro\meter}$-diameter bead.
The spectra were obtained with a $\SI{785}{\nano\meter}$-focused laser acting on a single bead by a $40\times$ objective lens with a power of \SI{2}{\milli\watt}.
Whereas the acoustofluidic device is set at \SI{1.056}{\mega\hertz} frequency and $\sim\SI{3.8}{\volt}$ voltage amplitude.
In all measurements, the exposure time is  \SI{30}{\second}, without accumulations. 
A notch filter smooths out the backscattered light from the undesirable elastic signal, e.g., the so-called Rayleigh scattering, which has the same wavelength as the excitation laser.
The inelastically scattered light passes towards the detection system to form the Raman spectrum.
The blue line in panels (a) and (c) present the spectrum taken at a particle in the levitation plane (referenced as $z=0$ in depth).
The obtained spectra are very alike and compatible with previously reported results.\cite{Sears1981}
In contrast, the red line corresponds to the spectrum at $z=-\SI{80}{\micro\meter}$ (underneath the levitation plane). 
As no particle is trapped at this height, the Raman signal is almost negligible.
To further investigate whether the single-particle spectrum in-depth, we laser-scanned along the axial line from $-60$ to $\SI{60}{\micro\meter}$.
Then we took 
the average Raman intensity around $\SI{1000}{\per\centi\meter}$ (e.g., the Raman signal is numerically integrated into a narrow interval around the peak intensity at $\SI{1000}{\per\centi\meter}$).
After fitting the obtained data with a Gaussian curve, we  estimate the full width at half maximum (FWHM) for the particles as $\text{FWHM}=\SI{15}{\micro\meter}$ (\SI{10}{\micro\meter} diameter) and $\text{FWHM}=\SI{33}{\micro\meter}$ (\SI{30}{\micro\meter} diameter).
This result is consistent with  focusing the Raman laser within a single-particle, albeit a deviation in diameter of  the smaller particle is noted.
We may attribute this discrepancy to a laser depth close to $\SI{2.0}{\micro\meter}$ and variations in the particle's size distribution. 
This result reasonably supports that a single layer of beads is formed at the levitation plane;
even considering higher concentrations in the range of $C=50$ to $900\,\textrm{particles/\si{\micro\liter}}$ (see Fig. S3\dag). 
For concentrations  as $C\ge\ 5000$\,particles/$\si{\micro\liter}$, the FWHM broadens due to the formation of other  microparticle layers underneath the levitation plane (see Fig. S4\dag).
\begin{figure}
    \begin{center}
        \includegraphics[scale=1]{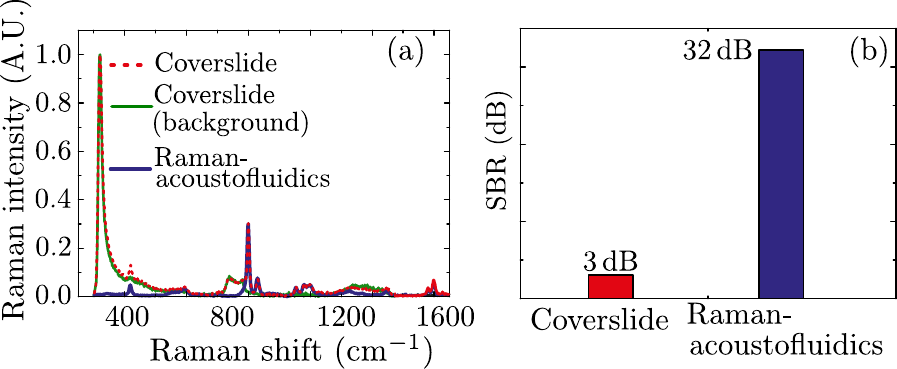}
        \caption{
            (a) The Raman spectrum of $\SI{30}{\micro\meter}$-diameter polystyrene beads for two distinct settings is plotted. 
            The red dotted line depicts the spectrum of sedimented particles on a silicon wafer (cover slide).
            Whereas the blue solid line represents the Raman spectrum of a single trapped particle.
            The background Raman signal (without particles) is also presented as the black solid line. 
            (b) The signal-to-background ratio (SBR) for each set is presented.
            The Raman-acoustofluidic platform shows an improvement of $\textrm{SBR}=\SI{32}{\decibel}$.
 \label{fig:Raman_comparison}
        }
    \end{center}
\end{figure}

Another aspect of the Raman-acoustofluidic platform
concerns the signal-to-background ratio (SBR). 
To analyse the SBR, we take the Raman spectrum of 
a $\SI{30}{\micro\meter}$-diameter particle in two distinct configurations. 
In one experiment, the spectrum is obtained in the Raman-acoustofludic system; while in the other,
we take the spectrum of particles 
sandwiched between a silicon wafer and a glass coverslip (conventional Raman method).
The obtained results 
are depicted in Fig.~\ref{fig:Raman_comparison}.
We note a remarkable difference in the results of  the two acquisition methods.
In the conventional setup (represented by the red dashed line),  a very large peak at 
\SI{550}{\per\centi\meter} is noticed, which corresponds to
the substrate (silicon wafer).
We also took the spectrum of the background only (the solid green line).
The solid blue line represents the spectrum acquired with the Raman-acoustofluidic settings.
Outstanding discrepancies is clearly noticed in the  $500$--$\SI{775}{\per\centi\meter}$, $915$--$\SI{990}{\per\centi\meter}$, and
$1275$--$\SI{1430}{\per\centi\meter}$ bands. 
Nearly three orders of magnitude enhancement are witnessed in SBR of the Raman-acoustofluidic platform regarding the conventional method. 
This result is depicted in dB, as shown in panel (b). 
The SBR definition  used in our calculation is given by
\begin{equation}
    \text{SBR}= 10\log\left(\frac{I_\text{T}- I_\text{B}}{I_\text{B}}\right).
\end{equation}
where $I_\text{T}$ and $I_\text{B}$ represent the total intensity signal and background intensity signal (sample is absent), respectively.

\subsection*{Macrophages of mice}
Let us now present our results considering biological cells. 
We have conducted experiments on macrophage cells of mice (cell line j774.A1), with an averaged diameter of $\SI{20}{\micro\meter}$.
The cells were immersed in phosphate buffer saline (PBS) solution and injected in the resonant chamber.
The cell trapping protocol is the same as that used for polystyrene beads.
The $\SI{785}{\nano\meter}$ laser was again employed because it excites the vibrational modes of the cells ($400$-$\SI{2000}{\per\centi\meter}$) efficiently.\cite{Wood2006}
Furthermore, as the Raman scattering efficiency of biological cells is much smaller than that of inorganic compounds,\cite{Kuhar2018, PUPPELS_1991} we had to increase the laser power from $2$ to $\SI{75}{\milli\watt}$.
Also, a longer acquisition time of $\SI{1}{\minute}$ with an average of $40$ accumulations was required to obtain the cellular spectrum.
Here no cell-ejection by light was observed. 

Raman-acoustofluidics may induce collateral damage due to the laser and acoustic interaction with cells, as well as the temperature rise and acoustic cavitation.
We have not observed photodamage in the probed cell after the Raman acquisition.
In fact, photodamage may not occur for a laser power as high as $\SI{115}{\milli\watt}$ for nearly one hour of cell exposure to light.\cite{Notingher2002}
Temperature rise may be caused by ultrasonic absorption in the liquid and biosample, heat losses in the piezoelectric actuator and in thin glue layers used to bond different device parts.\cite{Wiklund2012}
We have measured the overall chamber temperature with an infrared thermographic camera (E40bx, FLIR Systems, USA) with thermal sensitivity of $\SI{0.045}{\celsius}$ in the  $-20$ to $\SI{120}{\celsius}$ range.
At voltages as low as $\SI{8}{\volt}$, the temperature variations 
remained under $\SI{1}{\celsius}$ over a few hours measurement as shown in Fig. S5.\dag\:
Nonetheless, the local temperature of the Raman-investigated cell was not measured.
Appropriated approaches based on luminescent nanothermometers, which provide intracellular temperature readings,\cite{Jaque2012,Brites2016,Santos2018} can be employed for this task.
Also, the ultrasonic effects on cellular viability and metabolic activity were not investigated.
For reference, human dermal fibroblasts have good viability (above $80\,\%$) for a voltage range of $6$--$\SI{8}{\volt}$ and a time interval under $\SI{15}{\minute}$, when carried out at room temperature.\cite{Levario-Diaz2020}
Finally, a few words should be said about inertial cavitation that
may provoke cell lysis and death.
Inertial cavitation is caused by a large bubble oscillation over a few cycles ending in a violent bubble collapse. 
Since the cavitation threshold depends on the wave pressure amplitude, cavitation is less likely to take place in a pressure node.
Besides, for a frequency of $1$--$\SI{10}{\mega\hertz}$, the estimated pressure amplitude for cavitation is  $0.1$--$\SI{1}{\mega\pascal}$.\cite{Wiklund2012a}
This is well-above the estimated pressure amplitude inside the device chamber, e.g., $p_0=\SI{0.05}{\mega\pascal}$.
So it is unlike that inertial cavitation occurs in the levitation plane wherein cells are investigated.

\begin{figure}
    \begin{center}
        \includegraphics[scale=1.8]{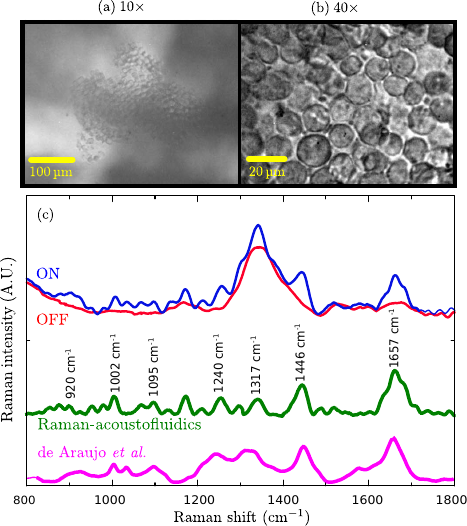}
        \caption{ 
        Bright-field images of macrophages in the levitation plane taken by an objective lens of (a) $10\times$ and (b) $40\times$. 
        The cell concentration is  $50$\,cells/\si{\micro\liter}. (c) The room-temperature Raman spectrum  of a single macrophage, with the device `off' (red curve) and `on' (blue curve), is obtained with the $40\times$ objective lens.
        The green curve is the Raman spectrum after background subtraction. 
        For comparison, the spectrum obtained from fixed macrophages\cite{Araujo2019} is also shown (pink curve).
 \label{fig:Raman_macrophage}
        }
    \end{center}
\end{figure}

In Fig.~\ref{fig:Raman_macrophage}, we show trapped macrophages and the  Raman fingerprint at room temperature of a single cell. 
We used a cell concentration of  $50$ cells/\si{\micro\liter}. 
The bright-field images in panels (a) and (b) show cells in the levitation plane after the Raman acquisition ($\sim \SI{1}{\hour}$).
The observed cells are well-packed in a stable configuration during the acquisition, without noticeable acoustic microstreaming effects.
By visual inspection, we see the membrane integrity and cell morphology are good indicators of cell preserved viability.
In panel (c), we exhibit the  Raman spectrum of a  macrophage.
The red and blue lines indicate the spectrum after switching the device `off' and `on,' respectively.
The differences between the spectra are notably straightforward. 
Indeed, before turning on the device, no Raman signal from a cell could be detected, and the only contribution comes from the culture medium (PBS). 
The green line represents the spectrum after subtracting the background signal.
This Raman signature is in accordance with previously reported results (pink curve) for the same cell line. \cite{Araujo2019} 
Several peaks in both spectra are alike.
However, some striking differences in the methodology of Ref.\cite{Araujo2019} should be pointed here.
They have used a $\SI{532}{\nano\meter}$ excitation laser focused through a $100\times$ objective lens ($\text{NA} = 0.9$); and  cells are fixed on a silicon wafer using a $4\%$-glutaraldehyde  solution (Merck KGaA, Germany).  
The spectrum delivered by Raman-acoustofluidics shows prominent peaks in the 
$900$--$\SI{1400}{\per\centi\meter}$ band, which are overshadowed in the result from  Ref.\cite{Araujo2019} 
Moreover, our system has a notable advantage over conventional fixing methods; it keeps cells alive during the entire acquisition process.

\section*{Conclusions}
We have presented a physical description of a Raman-acoustofluidic integrated platform for single-cell analysis.
The system comprises an acoustofluidic resonant chamber and a confocal Raman microscope.
The key aspects of the integrated system include particle trapping stability over a time interval compatible with the duration of cell assays under consideration, and a small temperature variation  ($\sim\SI{1}{\celsius}$).

The observed stable clumps of $\SI{10}{\micro\meter}$- and $\SI{30}{\micro\meter}$-diameter polystyrene beads in the levitation plane forming an hexagonal lattice appears to be the result of the secondary radiation force.
In this size range, acoustic microstreaming have not affected the trap stability specially for higher particle concentrations.
These results are achieved with an aggregation efficiency of $30$--$40\,\%$.  

The obtained Raman spectrum of a single polystyrene bead is in excellent agreement with previous observations.\cite{Sears1981}  
When a living macrophage is considered, the spectrum taken at room temperature shows a striking resemblance with the results obtained from fixed cells of the same line.\cite{Araujo2019}  
A better resolution in the Raman-acoustofluidic spectrum is nevertheless noted. 

Our study has expounded the Raman-acoustofluidics as a useful technique for single-cell analysis. 
Possible applications of the method span from drug discovery assays to tissue engineering investigations.

\section*{Author contributions}
This section describes the author contributions following the CRediT format. 
Harrisson D. A. Santos: Methodology, Investigation, Validation,  Writing - Original Draft. 
Amanda E. Silva:  Methodology, Investigation, Validation. 
Gicl\^enio C. Silva: Methodology, Software, Investigation, Validation. 
Everton B. Lima: Software, Validation.
Alisson S. Marques: Software, Validation.
Magna S. Alexandre-Moreira: Methodology, Resources, Supervision. 
Aline C. Queiroz: Methodology, Resources. 
Carlos Jacinto: Methodology, Resources. 
J. Henrique Lopes: Investigation, Resources. 
U\'eslen Rocha: Conceptualization, Methodology, Investigation, Validation.
Glauber T. Silva: Conceptualization,  Supervision, Writing - Review \& Editing, Project administration. 

\section*{Conflicts of interest}
There are no conflicts to declare.

\section*{Acknowledgements}
We acknowledge the financial support from Brazilian Agencies:
Funding Authority for Studies and Projects--FINEP (grants INFRAPESQ-11 and INFRAPESQ-12),
Council for Scientific and Technological Development--CNPq (grant numbers 431736/2018-9, 304967/2018-1, 308357/2019-1). H.D.A. Santos thanks the Federal University of Alagoas for postdoctoral scholarship.



\balance



\begin{mcitethebibliography}{60}
\providecommand*{\natexlab}[1]{#1}
\providecommand*{\mciteSetBstSublistMode}[1]{}
\providecommand*{\mciteSetBstMaxWidthForm}[2]{}
\providecommand*{\mciteBstWouldAddEndPuncttrue}
  {\def\EndOfBibitem{\unskip.}}
\providecommand*{\mciteBstWouldAddEndPunctfalse}
  {\let\EndOfBibitem\relax}
\providecommand*{\mciteSetBstMidEndSepPunct}[3]{}
\providecommand*{\mciteSetBstSublistLabelBeginEnd}[3]{}
\providecommand*{\EndOfBibitem}{}
\mciteSetBstSublistMode{f}
\mciteSetBstMaxWidthForm{subitem}
{(\emph{\alph{mcitesubitemcount}})}
\mciteSetBstSublistLabelBeginEnd{\mcitemaxwidthsubitemform\space}
{\relax}{\relax}

\bibitem[Seo \emph{et~al.}(2014)Seo, Kim, Joe, Han, Chen, Cheng, and
  Jang]{seo2014nir}
S.-H. Seo, B.-M. Kim, A.~Joe, H.-W. Han, X.~Chen, Z.~Cheng and E.-S. Jang,
  \emph{Biomaterials}, 2014, \textbf{35}, 3309--3318\relax
\mciteBstWouldAddEndPuncttrue
\mciteSetBstMidEndSepPunct{\mcitedefaultmidpunct}
{\mcitedefaultendpunct}{\mcitedefaultseppunct}\relax
\EndOfBibitem
\bibitem[Cui \emph{et~al.}(2018)Cui, Zhang, and Yue]{Cui2018}
S.~Cui, S.~Zhang and S.~Yue, \emph{Journal of Healthcare Engineering}, 2018,
  \textbf{2018}, 1--11\relax
\mciteBstWouldAddEndPuncttrue
\mciteSetBstMidEndSepPunct{\mcitedefaultmidpunct}
{\mcitedefaultendpunct}{\mcitedefaultseppunct}\relax
\EndOfBibitem
\bibitem[Taleb \emph{et~al.}(2006)Taleb, Diamond, McGarvey, Beattie, Toland,
  and Hamilton]{Taleb_2006}
A.~Taleb, J.~Diamond, J.~J. McGarvey, J.~R. Beattie, C.~Toland and P.~W.
  Hamilton, \emph{The Journal of Physical Chemistry B}, 2006, \textbf{110},
  19625--19631\relax
\mciteBstWouldAddEndPuncttrue
\mciteSetBstMidEndSepPunct{\mcitedefaultmidpunct}
{\mcitedefaultendpunct}{\mcitedefaultseppunct}\relax
\EndOfBibitem
\bibitem[Chan \emph{et~al.}(2009)Chan, Taylor, and Thompson]{Chan2009}
J.~W. Chan, D.~S. Taylor and D.~L. Thompson, \emph{Biopolymers}, 2009,
  \textbf{91}, 132--139\relax
\mciteBstWouldAddEndPuncttrue
\mciteSetBstMidEndSepPunct{\mcitedefaultmidpunct}
{\mcitedefaultendpunct}{\mcitedefaultseppunct}\relax
\EndOfBibitem
\bibitem[Notingher \emph{et~al.}(2004)Notingher, Jell, Lohbauer, Salih, and
  Hench]{Notingher2004}
I.~Notingher, G.~Jell, U.~Lohbauer, V.~Salih and L.~L. Hench, \emph{Journal of
  Cellular Biochemistry}, 2004, \textbf{92}, 1180--1192\relax
\mciteBstWouldAddEndPuncttrue
\mciteSetBstMidEndSepPunct{\mcitedefaultmidpunct}
{\mcitedefaultendpunct}{\mcitedefaultseppunct}\relax
\EndOfBibitem
\bibitem[Krishna \emph{et~al.}(2005)Krishna, Sockalingum, Kegelaer, Rubin,
  Kartha, and Manfait]{Krishna2005}
C.~M. Krishna, G.~D. Sockalingum, G.~Kegelaer, S.~Rubin, V.~B. Kartha and
  M.~Manfait, \emph{Vibrational Spectroscopy}, 2005, \textbf{38}, 95--100\relax
\mciteBstWouldAddEndPuncttrue
\mciteSetBstMidEndSepPunct{\mcitedefaultmidpunct}
{\mcitedefaultendpunct}{\mcitedefaultseppunct}\relax
\EndOfBibitem
\bibitem[Chan \emph{et~al.}(2006)Chan, Taylor, Zwerdling, Lane, Ihara, and
  Huser]{Chan2006}
J.~W. Chan, D.~S. Taylor, T.~Zwerdling, S.~M. Lane, K.~Ihara and T.~Huser,
  \emph{Biophysical Journal}, 2006, \textbf{90}, 648--656\relax
\mciteBstWouldAddEndPuncttrue
\mciteSetBstMidEndSepPunct{\mcitedefaultmidpunct}
{\mcitedefaultendpunct}{\mcitedefaultseppunct}\relax
\EndOfBibitem
\bibitem[Frank \emph{et~al.}(1995)Frank, McCreery, and Redd]{Frank1995}
C.~J. Frank, R.~L. McCreery and D.~C.~B. Redd, \emph{Analytical Chemistry},
  1995, \textbf{67}, 777--783\relax
\mciteBstWouldAddEndPuncttrue
\mciteSetBstMidEndSepPunct{\mcitedefaultmidpunct}
{\mcitedefaultendpunct}{\mcitedefaultseppunct}\relax
\EndOfBibitem
\bibitem[Krishna \emph{et~al.}(2004)Krishna, Sockalingum, Kurien, Rao, Venteo,
  Pluot, Manfait, and Kartha]{Krishna2004}
C.~M. Krishna, G.~D. Sockalingum, J.~Kurien, L.~Rao, L.~Venteo, M.~Pluot,
  M.~Manfait and V.~B. Kartha, \emph{Applied Spectroscopy}, 2004, \textbf{58},
  1128--1135\relax
\mciteBstWouldAddEndPuncttrue
\mciteSetBstMidEndSepPunct{\mcitedefaultmidpunct}
{\mcitedefaultendpunct}{\mcitedefaultseppunct}\relax
\EndOfBibitem
\bibitem[Short \emph{et~al.}(2005)Short, Carpenter, Freyer, and
  Mourant]{Short2005}
K.~W. Short, S.~Carpenter, J.~P. Freyer and J.~R. Mourant, \emph{Biophysical
  Journal}, 2005, \textbf{88}, 4274--4288\relax
\mciteBstWouldAddEndPuncttrue
\mciteSetBstMidEndSepPunct{\mcitedefaultmidpunct}
{\mcitedefaultendpunct}{\mcitedefaultseppunct}\relax
\EndOfBibitem
\bibitem[Alraies \emph{et~al.}(2019)Alraies, Canetta, Waddington, Moseley, and
  Sloan]{Alraies2019}
A.~Alraies, E.~Canetta, R.~J. Waddington, R.~Moseley and A.~J. Sloan,
  \emph{Tissue Engineering Part C: Methods}, 2019, \textbf{25}, 489--499\relax
\mciteBstWouldAddEndPuncttrue
\mciteSetBstMidEndSepPunct{\mcitedefaultmidpunct}
{\mcitedefaultendpunct}{\mcitedefaultseppunct}\relax
\EndOfBibitem
\bibitem[{\v{Z}}ivanovi{\'{c}} \emph{et~al.}(2018){\v{Z}}ivanovi{\'{c}},
  Semini, Laue, Drescher, Aebischer, and Kneipp]{Zivanovic2018}
V.~{\v{Z}}ivanovi{\'{c}}, G.~Semini, M.~Laue, D.~Drescher, T.~Aebischer and
  J.~Kneipp, \emph{Analytical Chemistry}, 2018, \textbf{90}, 8154--8161\relax
\mciteBstWouldAddEndPuncttrue
\mciteSetBstMidEndSepPunct{\mcitedefaultmidpunct}
{\mcitedefaultendpunct}{\mcitedefaultseppunct}\relax
\EndOfBibitem
\bibitem[Bik \emph{et~al.}(2019)Bik, Mielniczek, Jarosz, Denbigh, Budzynska,
  Baranska, and Majzner]{Bik2019}
E.~Bik, N.~Mielniczek, M.~Jarosz, J.~Denbigh, R.~Budzynska, M.~Baranska and
  K.~Majzner, \emph{The Analyst}, 2019, \textbf{144}, 6561--6569\relax
\mciteBstWouldAddEndPuncttrue
\mciteSetBstMidEndSepPunct{\mcitedefaultmidpunct}
{\mcitedefaultendpunct}{\mcitedefaultseppunct}\relax
\EndOfBibitem
\bibitem[Nguyen \emph{et~al.}(2017)Nguyen, Peters, and Schultz]{Nguyen2017}
A.~H. Nguyen, E.~A. Peters and Z.~D. Schultz, \emph{Reviews in Analytical
  Chemistry}, 2017, \textbf{36}, year\relax
\mciteBstWouldAddEndPuncttrue
\mciteSetBstMidEndSepPunct{\mcitedefaultmidpunct}
{\mcitedefaultendpunct}{\mcitedefaultseppunct}\relax
\EndOfBibitem
\bibitem[Li \emph{et~al.}(2014)Li, Deen, Kumar, and Selvaganapathy]{Li2014}
Z.~Li, M.~Deen, S.~Kumar and P.~Selvaganapathy, \emph{Sensors}, 2014,
  \textbf{14}, 17275--17303\relax
\mciteBstWouldAddEndPuncttrue
\mciteSetBstMidEndSepPunct{\mcitedefaultmidpunct}
{\mcitedefaultendpunct}{\mcitedefaultseppunct}\relax
\EndOfBibitem
\bibitem[Zhao \emph{et~al.}(2008)Zhao, Lui, McLean, and Zeng]{Zhao2008}
J.~Zhao, H.~Lui, D.~I. McLean and H.~Zeng, \emph{Skin Research and Technology},
  2008, \textbf{14}, 484--492\relax
\mciteBstWouldAddEndPuncttrue
\mciteSetBstMidEndSepPunct{\mcitedefaultmidpunct}
{\mcitedefaultendpunct}{\mcitedefaultseppunct}\relax
\EndOfBibitem
\bibitem[Jones \emph{et~al.}(2019)Jones, Hooper, Zhang, Wolverson, and
  Valev]{Jones2019}
R.~R. Jones, D.~C. Hooper, L.~Zhang, D.~Wolverson and V.~K. Valev,
  \emph{Nanoscale Research Letters}, 2019, \textbf{14}, year\relax
\mciteBstWouldAddEndPuncttrue
\mciteSetBstMidEndSepPunct{\mcitedefaultmidpunct}
{\mcitedefaultendpunct}{\mcitedefaultseppunct}\relax
\EndOfBibitem
\bibitem[Puppels(1991)]{PUPPELS_1991}
G.~Puppels, \emph{Experimental Cell Research}, 1991, \textbf{195},
  361--367\relax
\mciteBstWouldAddEndPuncttrue
\mciteSetBstMidEndSepPunct{\mcitedefaultmidpunct}
{\mcitedefaultendpunct}{\mcitedefaultseppunct}\relax
\EndOfBibitem
\bibitem[Mason and Leary(1991)]{Mason1991}
J.~T. Mason and T.~J.~O. Leary, \emph{Journal of Histochemistry {\&}
  Cytochemistry}, 1991, \textbf{39}, 225--229\relax
\mciteBstWouldAddEndPuncttrue
\mciteSetBstMidEndSepPunct{\mcitedefaultmidpunct}
{\mcitedefaultendpunct}{\mcitedefaultseppunct}\relax
\EndOfBibitem
\bibitem[Li \emph{et~al.}(2017)Li, Almassalha, Chandler, Zhou, Stypula-Cyrus,
  Hujsak, Roth, Bleher, Subramanian, Szleifer, Dravid, and Backman]{Li2017}
Y.~Li, L.~M. Almassalha, J.~E. Chandler, X.~Zhou, Y.~E. Stypula-Cyrus, K.~A.
  Hujsak, E.~W. Roth, R.~Bleher, H.~Subramanian, I.~Szleifer, V.~P. Dravid and
  V.~Backman, \emph{Experimental Cell Research}, 2017, \textbf{358},
  253--259\relax
\mciteBstWouldAddEndPuncttrue
\mciteSetBstMidEndSepPunct{\mcitedefaultmidpunct}
{\mcitedefaultendpunct}{\mcitedefaultseppunct}\relax
\EndOfBibitem
\bibitem[Snook \emph{et~al.}(2009)Snook, Harvey, Faria, and Gardner]{Snook2009}
R.~D. Snook, T.~J. Harvey, E.~C. Faria and P.~Gardner, \emph{Integr. Biol.},
  2009, \textbf{1}, 43--52\relax
\mciteBstWouldAddEndPuncttrue
\mciteSetBstMidEndSepPunct{\mcitedefaultmidpunct}
{\mcitedefaultendpunct}{\mcitedefaultseppunct}\relax
\EndOfBibitem
\bibitem[Bl{\'{a}}zquez-Castro(2019)]{BlazquezCastro2019}
A.~Bl{\'{a}}zquez-Castro, \emph{Micromachines}, 2019, \textbf{10}, 507\relax
\mciteBstWouldAddEndPuncttrue
\mciteSetBstMidEndSepPunct{\mcitedefaultmidpunct}
{\mcitedefaultendpunct}{\mcitedefaultseppunct}\relax
\EndOfBibitem
\bibitem[Dholakia \emph{et~al.}(2020)Dholakia, Drinkwater, and
  Ritsch-Marte]{Dholakia2020}
K.~Dholakia, B.~W. Drinkwater and M.~Ritsch-Marte, \emph{Nature Reviews
  Physics}, 2020, \textbf{2}, 480--491\relax
\mciteBstWouldAddEndPuncttrue
\mciteSetBstMidEndSepPunct{\mcitedefaultmidpunct}
{\mcitedefaultendpunct}{\mcitedefaultseppunct}\relax
\EndOfBibitem
\bibitem[Bruus \emph{et~al.}(2011)Bruus, Dual, Hawkes, Hill, Laurell, Nilsson,
  Radel, Sadhal, and Wiklund]{Bruus2011}
H.~Bruus, J.~Dual, J.~Hawkes, M.~Hill, T.~Laurell, J.~Nilsson, S.~Radel,
  S.~Sadhal and M.~Wiklund, \emph{Lab Chip}, 2011, \textbf{11},
  3579--3580\relax
\mciteBstWouldAddEndPuncttrue
\mciteSetBstMidEndSepPunct{\mcitedefaultmidpunct}
{\mcitedefaultendpunct}{\mcitedefaultseppunct}\relax
\EndOfBibitem
\bibitem[Spengler \emph{et~al.}(2001)Spengler, Jekel, Christensen, Adrian,
  Hawkes, and Coakley]{Spengler2001}
J.~F. Spengler, M.~Jekel, K.~T. Christensen, R.~J. Adrian, J.~J. Hawkes and
  W.~T. Coakley, \emph{Bioseparation}, 2001, \textbf{9}, 329--341\relax
\mciteBstWouldAddEndPuncttrue
\mciteSetBstMidEndSepPunct{\mcitedefaultmidpunct}
{\mcitedefaultendpunct}{\mcitedefaultseppunct}\relax
\EndOfBibitem
\bibitem[Christakou \emph{et~al.}(2015)Christakou, Mathias~Ohlin, and
  Wiklund]{Christakou2015}
A.~E. Christakou, B.~O. Mathias~Ohlin and M.~Wiklund, \emph{Lab Chip}, 2015,
  \textbf{15}, 3222--3231\relax
\mciteBstWouldAddEndPuncttrue
\mciteSetBstMidEndSepPunct{\mcitedefaultmidpunct}
{\mcitedefaultendpunct}{\mcitedefaultseppunct}\relax
\EndOfBibitem
\bibitem[Biswas(1995)]{Biswas1995}
A.~Biswas, \emph{Journal of Crystal Growth}, 1995, \textbf{147}, 155--164\relax
\mciteBstWouldAddEndPuncttrue
\mciteSetBstMidEndSepPunct{\mcitedefaultmidpunct}
{\mcitedefaultendpunct}{\mcitedefaultseppunct}\relax
\EndOfBibitem
\bibitem[Santesson \emph{et~al.}(2003)Santesson, Johansson, Taylor, Levander,
  Fox, Sepaniak, and Nilsson]{Santesson2003}
S.~Santesson, J.~Johansson, L.~S. Taylor, I.~Levander, S.~Fox, M.~Sepaniak and
  S.~Nilsson, \emph{Analytical Chemistry}, 2003, \textbf{75}, 2177--2180\relax
\mciteBstWouldAddEndPuncttrue
\mciteSetBstMidEndSepPunct{\mcitedefaultmidpunct}
{\mcitedefaultendpunct}{\mcitedefaultseppunct}\relax
\EndOfBibitem
\bibitem[Leopold \emph{et~al.}(2003)Leopold, Haberkorn, Laurell, Nilsson,
  Baena, Frank, and Lendl]{Leopold2003}
N.~Leopold, M.~Haberkorn, T.~Laurell, J.~Nilsson, J.~R. Baena, J.~Frank and
  B.~Lendl, \emph{Analytical Chemistry}, 2003, \textbf{75}, 2166--2171\relax
\mciteBstWouldAddEndPuncttrue
\mciteSetBstMidEndSepPunct{\mcitedefaultmidpunct}
{\mcitedefaultendpunct}{\mcitedefaultseppunct}\relax
\EndOfBibitem
\bibitem[L{\'{o}}pez-Pastor \emph{et~al.}(2007)L{\'{o}}pez-Pastor,
  Dom{\'{\i}}nguez-Vidal, Ayora-Ca{\~{n}}ada, Laurell, Valc{\'{a}}rcel, and
  Lendl]{LopezPastor2007}
M.~L{\'{o}}pez-Pastor, A.~Dom{\'{\i}}nguez-Vidal, M.~J. Ayora-Ca{\~{n}}ada,
  T.~Laurell, M.~Valc{\'{a}}rcel and B.~Lendl, \emph{Lab Chip}, 2007,
  \textbf{7}, 126--132\relax
\mciteBstWouldAddEndPuncttrue
\mciteSetBstMidEndSepPunct{\mcitedefaultmidpunct}
{\mcitedefaultendpunct}{\mcitedefaultseppunct}\relax
\EndOfBibitem
\bibitem[Qui{\~{n}}o \emph{et~al.}(2015)Qui{\~{n}}o, Hellwig, Griesing, Pauer,
  Moritz, Will, and Braeuer]{Quino2015}
J.~Qui{\~{n}}o, T.~Hellwig, M.~Griesing, W.~Pauer, H.-U. Moritz, S.~Will and
  A.~Braeuer, \emph{International Journal of Heat and Mass Transfer}, 2015,
  \textbf{89}, 406--413\relax
\mciteBstWouldAddEndPuncttrue
\mciteSetBstMidEndSepPunct{\mcitedefaultmidpunct}
{\mcitedefaultendpunct}{\mcitedefaultseppunct}\relax
\EndOfBibitem
\bibitem[Wood \emph{et~al.}(2005)Wood, Heraud, Stojkovic, Morrison, Beardall,
  and McNaughton]{Wood2005}
B.~R. Wood, P.~Heraud, S.~Stojkovic, D.~Morrison, J.~Beardall and
  D.~McNaughton, \emph{Analytical Chemistry}, 2005, \textbf{77},
  4955--4961\relax
\mciteBstWouldAddEndPuncttrue
\mciteSetBstMidEndSepPunct{\mcitedefaultmidpunct}
{\mcitedefaultendpunct}{\mcitedefaultseppunct}\relax
\EndOfBibitem
\bibitem[Puskar \emph{et~al.}(2007)Puskar, Tuckermann, Frosch, Popp, Ly,
  McNaughton, and Wood]{Puskar2007}
L.~Puskar, R.~Tuckermann, T.~Frosch, J.~Popp, V.~Ly, D.~McNaughton and B.~R.
  Wood, \emph{Lab on a Chip}, 2007, \textbf{7}, 1125\relax
\mciteBstWouldAddEndPuncttrue
\mciteSetBstMidEndSepPunct{\mcitedefaultmidpunct}
{\mcitedefaultendpunct}{\mcitedefaultseppunct}\relax
\EndOfBibitem
\bibitem[Ruedas-Rama \emph{et~al.}(2007)Ruedas-Rama, Dom{\'{\i}}nguez-Vidal,
  Radel, and Lendl]{RuedasRama2007}
M.~J. Ruedas-Rama, A.~Dom{\'{\i}}nguez-Vidal, S.~Radel and B.~Lendl,
  \emph{Analytical Chemistry}, 2007, \textbf{79}, 7853--7857\relax
\mciteBstWouldAddEndPuncttrue
\mciteSetBstMidEndSepPunct{\mcitedefaultmidpunct}
{\mcitedefaultendpunct}{\mcitedefaultseppunct}\relax
\EndOfBibitem
\bibitem[Wieland \emph{et~al.}(2019)Wieland, Tauber, Gasser, Rettenbacher, Lux,
  Radel, and Lendl]{Wieland2019}
K.~Wieland, S.~Tauber, C.~Gasser, L.~A. Rettenbacher, L.~Lux, S.~Radel and
  B.~Lendl, \emph{Analytical Chemistry}, 2019, \textbf{91}, 14231--14238\relax
\mciteBstWouldAddEndPuncttrue
\mciteSetBstMidEndSepPunct{\mcitedefaultmidpunct}
{\mcitedefaultendpunct}{\mcitedefaultseppunct}\relax
\EndOfBibitem
\bibitem[Baron \emph{et~al.}(2020)Baron, Chen, Hammarstrom, Hammond,
  Glynne-Jones, Gillespie, and Dholakia]{Baron2020}
V.~O. Baron, M.~Chen, B.~Hammarstrom, R.~J.~H. Hammond, P.~Glynne-Jones, S.~H.
  Gillespie and K.~Dholakia, \emph{Communications Biology}, 2020, \textbf{3},
  236\relax
\mciteBstWouldAddEndPuncttrue
\mciteSetBstMidEndSepPunct{\mcitedefaultmidpunct}
{\mcitedefaultendpunct}{\mcitedefaultseppunct}\relax
\EndOfBibitem
\bibitem[Xu \emph{et~al.}(2020)Xu, Luo, Liu, Zhang, and Wang]{Xu2020}
T.~Xu, Y.~Luo, C.~Liu, X.~Zhang and S.~Wang, \emph{Analytical Chemistry}, 2020,
  \textbf{92}, 7816--7821\relax
\mciteBstWouldAddEndPuncttrue
\mciteSetBstMidEndSepPunct{\mcitedefaultmidpunct}
{\mcitedefaultendpunct}{\mcitedefaultseppunct}\relax
\EndOfBibitem
\bibitem[Corle and Kino(1996)]{Corle1996}
T.~R. Corle and G.~S. Kino, \emph{Confocal Scanning Optical Microscopy and
  Related Imaging Systems}, Academic Press, Burlington, 1996, pp. 147 --
  223\relax
\mciteBstWouldAddEndPuncttrue
\mciteSetBstMidEndSepPunct{\mcitedefaultmidpunct}
{\mcitedefaultendpunct}{\mcitedefaultseppunct}\relax
\EndOfBibitem
\bibitem[Ara{\'{u}}jo \emph{et~al.}(2019)Ara{\'{u}}jo, Queiroz, Silva, Silva,
  Silva, Silva, Silva, Souza, Fonseca, Camara, Silva, and
  Alexandre-Moreira]{Araujo2019}
M.~V. Ara{\'{u}}jo, A.~C. Queiroz, J.~F.~M. Silva, A.~E. Silva, J.~K.~S. Silva,
  G.~R. Silva, E.~C.~O. Silva, S.~T. Souza, E.~J.~S. Fonseca, C.~A. Camara,
  T.~M.~S. Silva and M.~S. Alexandre-Moreira, \emph{The Analyst}, 2019,
  \textbf{144}, 5232--5244\relax
\mciteBstWouldAddEndPuncttrue
\mciteSetBstMidEndSepPunct{\mcitedefaultmidpunct}
{\mcitedefaultendpunct}{\mcitedefaultseppunct}\relax
\EndOfBibitem
\bibitem[Bruus(2012)]{Bruus2012}
H.~Bruus, \emph{Lab on a Chip}, 2012, \textbf{12}, 1014\relax
\mciteBstWouldAddEndPuncttrue
\mciteSetBstMidEndSepPunct{\mcitedefaultmidpunct}
{\mcitedefaultendpunct}{\mcitedefaultseppunct}\relax
\EndOfBibitem
\bibitem[Silva(2014)]{Silva2014}
G.~T. Silva, \emph{The Journal of the Acoustical Society of America}, 2014,
  \textbf{136}, 2405--2413\relax
\mciteBstWouldAddEndPuncttrue
\mciteSetBstMidEndSepPunct{\mcitedefaultmidpunct}
{\mcitedefaultendpunct}{\mcitedefaultseppunct}\relax
\EndOfBibitem
\bibitem[Wiklund(2012)]{Wiklund2012a}
M.~Wiklund, \emph{Lab Chip}, 2012, \textbf{12}, 2018--2028\relax
\mciteBstWouldAddEndPuncttrue
\mciteSetBstMidEndSepPunct{\mcitedefaultmidpunct}
{\mcitedefaultendpunct}{\mcitedefaultseppunct}\relax
\EndOfBibitem
\bibitem[Silva and Bruus(2014)]{Silva2014a}
G.~T. Silva and H.~Bruus, \emph{Physical Review E}, 2014, \textbf{90},
  063007\relax
\mciteBstWouldAddEndPuncttrue
\mciteSetBstMidEndSepPunct{\mcitedefaultmidpunct}
{\mcitedefaultendpunct}{\mcitedefaultseppunct}\relax
\EndOfBibitem
\bibitem[Lopes \emph{et~al.}(2016)Lopes, Azarpeyvand, and Silva]{Lopes2016}
J.~H. Lopes, M.~Azarpeyvand and G.~T. Silva, \emph{{IEEE} Transactions on
  Ultrasonics, Ferroelectrics, and Frequency Control}, 2016, \textbf{63},
  186--197\relax
\mciteBstWouldAddEndPuncttrue
\mciteSetBstMidEndSepPunct{\mcitedefaultmidpunct}
{\mcitedefaultendpunct}{\mcitedefaultseppunct}\relax
\EndOfBibitem
\bibitem[Pierce(2019)]{pierce2019acoustics}
A.~D. Pierce, \emph{Acoustics: an introduction to its physical principles and
  applications}, Springer, 2019\relax
\mciteBstWouldAddEndPuncttrue
\mciteSetBstMidEndSepPunct{\mcitedefaultmidpunct}
{\mcitedefaultendpunct}{\mcitedefaultseppunct}\relax
\EndOfBibitem
\bibitem[Wiklund \emph{et~al.}(2012)Wiklund, Green, and Ohlin]{Wiklund2012}
M.~Wiklund, R.~Green and M.~Ohlin, \emph{Lab on a Chip}, 2012, \textbf{12},
  2438\relax
\mciteBstWouldAddEndPuncttrue
\mciteSetBstMidEndSepPunct{\mcitedefaultmidpunct}
{\mcitedefaultendpunct}{\mcitedefaultseppunct}\relax
\EndOfBibitem
\bibitem[Spengler \emph{et~al.}(2003)Spengler, Coakley, and
  Christensen]{Spengler2003}
J.~F. Spengler, W.~T. Coakley and K.~T. Christensen, \emph{{AIChE} Journal},
  2003, \textbf{49}, 2773--2782\relax
\mciteBstWouldAddEndPuncttrue
\mciteSetBstMidEndSepPunct{\mcitedefaultmidpunct}
{\mcitedefaultendpunct}{\mcitedefaultseppunct}\relax
\EndOfBibitem
\bibitem[Bach and Bruus(2020)]{Bach2020}
J.~S. Bach and H.~Bruus, \emph{Physical Review Letters}, 2020, \textbf{124},
  year\relax
\mciteBstWouldAddEndPuncttrue
\mciteSetBstMidEndSepPunct{\mcitedefaultmidpunct}
{\mcitedefaultendpunct}{\mcitedefaultseppunct}\relax
\EndOfBibitem
\bibitem[Collins \emph{et~al.}(2015)Collins, Morahan, Garcia-Bustos, Doerig,
  Plebanski, and Neild]{Collins2015}
D.~J. Collins, B.~Morahan, J.~Garcia-Bustos, C.~Doerig, M.~Plebanski and
  A.~Neild, \emph{Nat Comm}, 2015, \textbf{6}, 8686\relax
\mciteBstWouldAddEndPuncttrue
\mciteSetBstMidEndSepPunct{\mcitedefaultmidpunct}
{\mcitedefaultendpunct}{\mcitedefaultseppunct}\relax
\EndOfBibitem
\bibitem[Silva \emph{et~al.}(2019)Silva, Lopes, ao~Neto, Nichols, and
  Drinkwater]{Silva2019}
G.~T. Silva, J.~H. Lopes, J.~P.~L. ao~Neto, M.~K. Nichols and B.~W. Drinkwater,
  \emph{Phys Rev Appl}, 2019, \textbf{11}, 054044\relax
\mciteBstWouldAddEndPuncttrue
\mciteSetBstMidEndSepPunct{\mcitedefaultmidpunct}
{\mcitedefaultendpunct}{\mcitedefaultseppunct}\relax
\EndOfBibitem
\bibitem[Campello and Cassares(2016)]{Campello2016}
E.~M.~B. Campello and K.~R. Cassares, \emph{Latin American Journal of Solids
  and Structures}, 2016, \textbf{13}, 23--50\relax
\mciteBstWouldAddEndPuncttrue
\mciteSetBstMidEndSepPunct{\mcitedefaultmidpunct}
{\mcitedefaultendpunct}{\mcitedefaultseppunct}\relax
\EndOfBibitem
\bibitem[Dumy \emph{et~al.}(2019)Dumy, Hoyos, and Aider]{Dumy2019}
G.~Dumy, M.~Hoyos and J.-L. Aider, \emph{The Journal of the Acoustical Society
  of America}, 2019, \textbf{146}, 4557--4568\relax
\mciteBstWouldAddEndPuncttrue
\mciteSetBstMidEndSepPunct{\mcitedefaultmidpunct}
{\mcitedefaultendpunct}{\mcitedefaultseppunct}\relax
\EndOfBibitem
\bibitem[Sears \emph{et~al.}(1981)Sears, Hunt, and Stevens]{Sears1981}
W.~M. Sears, J.~L. Hunt and J.~R. Stevens, \emph{The Journal of Chemical
  Physics}, 1981, \textbf{75}, 1589--1598\relax
\mciteBstWouldAddEndPuncttrue
\mciteSetBstMidEndSepPunct{\mcitedefaultmidpunct}
{\mcitedefaultendpunct}{\mcitedefaultseppunct}\relax
\EndOfBibitem
\bibitem[Wood \emph{et~al.}(2006)Wood, Caspers, Puppels, Pandiancherri, and
  McNaughton]{Wood2006}
B.~R. Wood, P.~Caspers, G.~J. Puppels, S.~Pandiancherri and D.~McNaughton,
  \emph{Analytical and Bioanalytical Chemistry}, 2006, \textbf{387},
  1691--1703\relax
\mciteBstWouldAddEndPuncttrue
\mciteSetBstMidEndSepPunct{\mcitedefaultmidpunct}
{\mcitedefaultendpunct}{\mcitedefaultseppunct}\relax
\EndOfBibitem
\bibitem[Kuhar \emph{et~al.}(2018)Kuhar, Sil, Verma, and Umapathy]{Kuhar2018}
N.~Kuhar, S.~Sil, T.~Verma and S.~Umapathy, \emph{{RSC} Advances}, 2018,
  \textbf{8}, 25888--25908\relax
\mciteBstWouldAddEndPuncttrue
\mciteSetBstMidEndSepPunct{\mcitedefaultmidpunct}
{\mcitedefaultendpunct}{\mcitedefaultseppunct}\relax
\EndOfBibitem
\bibitem[Notingher \emph{et~al.}(2002)Notingher, Verrier, Romanska, Bishop,
  Polak, and Hench]{Notingher2002}
I.~Notingher, S.~Verrier, H.~Romanska, A.~E. Bishop, J.~M. Polak and L.~L.
  Hench, \emph{Spectroscopy}, 2002, \textbf{16}, 43--51\relax
\mciteBstWouldAddEndPuncttrue
\mciteSetBstMidEndSepPunct{\mcitedefaultmidpunct}
{\mcitedefaultendpunct}{\mcitedefaultseppunct}\relax
\EndOfBibitem
\bibitem[Jaque and Vetrone(2012)]{Jaque2012}
D.~Jaque and F.~Vetrone, \emph{Nanoscale}, 2012, \textbf{4}, 4301\relax
\mciteBstWouldAddEndPuncttrue
\mciteSetBstMidEndSepPunct{\mcitedefaultmidpunct}
{\mcitedefaultendpunct}{\mcitedefaultseppunct}\relax
\EndOfBibitem
\bibitem[Brites \emph{et~al.}(2016)Brites, Mill{\'{a}}n, and
  Carlos]{Brites2016}
C.~Brites, A.~Mill{\'{a}}n and L.~Carlos, \emph{Including Actinides}, Elsevier,
  2016, pp. 339--427\relax
\mciteBstWouldAddEndPuncttrue
\mciteSetBstMidEndSepPunct{\mcitedefaultmidpunct}
{\mcitedefaultendpunct}{\mcitedefaultseppunct}\relax
\EndOfBibitem
\bibitem[Santos \emph{et~al.}(2018)Santos, Ximendes, del Carmen Iglesias-de~la
  Cruz, Chaves-Coira, del Rosal, Jacinto, Monge, Rubia-Rodr{\'{\i}}guez,
  Ortega, Mateos, Garc{\'{\i}}aSol{\'{e}}, Jaque, and
  Fern{\'{a}}ndez]{Santos2018}
H.~D.~A. Santos, E.~C. Ximendes, M.~del Carmen Iglesias-de~la Cruz,
  I.~Chaves-Coira, B.~del Rosal, C.~Jacinto, L.~Monge,
  I.~Rubia-Rodr{\'{\i}}guez, D.~Ortega, S.~Mateos, J.~Garc{\'{\i}}aSol{\'{e}},
  D.~Jaque and N.~Fern{\'{a}}ndez, \emph{Advanced Functional Materials}, 2018,
  \textbf{28}, 1803924\relax
\mciteBstWouldAddEndPuncttrue
\mciteSetBstMidEndSepPunct{\mcitedefaultmidpunct}
{\mcitedefaultendpunct}{\mcitedefaultseppunct}\relax
\EndOfBibitem
\bibitem[Victoria Levario-Diaz \emph{et~al.}(2020)Victoria Levario-Diaz, Galan,
  and Barnes]{Levario-Diaz2020}
P.~B. Victoria Levario-Diaz, M.~C. Galan and A.~C. Barnes, \emph{Sci Rep},
  2020, \textbf{10}, 8493\relax
\mciteBstWouldAddEndPuncttrue
\mciteSetBstMidEndSepPunct{\mcitedefaultmidpunct}
{\mcitedefaultendpunct}{\mcitedefaultseppunct}\relax
\EndOfBibitem
\end{mcitethebibliography}

\providecommand*{\mcitethebibliography}{\thebibliography}
\csname @ifundefined\endcsname{endmcitethebibliography}
{\let\endmcitethebibliography\endthebibliography}{}

\end{document}